\documentclass[11pt]{article}

\usepackage[utf8]{inputenc}
\usepackage{authblk}
\usepackage{graphicx}
\usepackage{amsmath, amssymb}
\usepackage[breaklinks=true]{hyperref}
\usepackage{xurl} 
\usepackage{csquotes}
\usepackage{xcolor}
\usepackage[margin=3cm]{geometry}
\usepackage{comment}
\hypersetup{
  colorlinks=true,
  linkcolor=black,
  urlcolor=blue
}
\title{Strengthening Energy Access in Remote Off-Grid Contexts: From Deployment to Long-Term System Availability}

\author[1]{Sveva Sodomaco}
\author[2]{Giacomo Sofi}
\author[2]{Mattia Erroi}
\author[2]{Niccolò Manfredi Selvaggi}
\author[2]{Gabriele Schiffo}
\author[2]{Rubens Dellepiane}

\affil[1]{Scuola Normale Superiore, Pisa, Italy}
\affil[2]{Collège des Ingénieurs, Paris, France}

\date{}

\begin{document}

\maketitle

\begin{center}
E-mail: \textit{sveva.sodomaco@sns.it, giacomo.sofi@cdi.eu}
\end{center}
\vspace{1em}

\begin{abstract}
Decentralized renewable energy (DRE) systems have become a cornerstone of electrification efforts in remote and underserved areas. Yet, while global attention has focused on expanding access through solar mini-grids and off-grid solutions, far less emphasis has been placed on ensuring the long-term operation of these systems. In many fragile contexts, weak Operation and Maintenance (O\&M) frameworks undermine the reliability and resilience of energy access, leading to premature system failure. This Perspective examines how structural and contextual barriers such as limited technical expertise, inadequate maintenance models, and insufficient local integration of energy systems contribute to this hidden challenge. We argue that O\&M should be reframed as a strategic function and embedded from the design phase onward. Building on insights from energy-scarce regions, we outline a pathway that combines user-friendly maintenance protocols, offline educational platforms, and community-based toolkits to support energy availability. This approach aims to empower local actors, support system functionality, and advance a sustainable energy transition.
\end{abstract}

\begin{center}
\end{center}

\section{Introduction}
The global energy system is undergoing a profound transformation, driven by climate change, technological innovation, and socio-economic shifts. By 2050, the world population is expected to reach 10 billion, significantly increasing the demand for energy services such as lighting, mobility, heating, and industry.\cite{IRENA2024} Although efficiency gains and electrification will moderate this growth, especially in high-income regions, global electricity demand is projected to increase sharply, potentially exceeding 50\% of the final energy consumption by 2050.\cite{IRENA2024,IEA2024} This trend of electrification is accompanied by a transition in energy supply systems. Renewables, which are essential for decarbonization and now often cheaper than fossil fuels even without accounting for externalities, are expected to dominate future energy mix.\cite{REF2022} Electricity networks, in particular, will face growing shares of variable renewable energy, which, being nondispatchable, require major investments in storage and grid flexibility (e.g. batteries, hydrogen, vehicle-to-grid technologies).\cite{IPCC2011}

The combination of increasing energy demand and the growing penetration of renewable energy sources is driving a paradigm shift from centralized to decentralized energy systems. These systems enhance energy security, reduce transmission losses, and empower communities to manage their own energy needs.\cite{Euroheat2023} Decentralized energy systems are also crucial for climate adaptation, as they enable faster recovery from climate-induced grid disruptions.However, while decentralization enhances system flexibility and resilience, it often delegates operational responsibility to local stakeholders. In remote and underserved contexts, which are characterized by geographic isolation, limited infrastructure, and weak institutional support, this shift introduces new challenges.\cite{Energypedia2025} Without adequate Operation and Maintenance (O\&M) frameworks, even properly designed renewable energy systems may suffer premature failure, as documented in recent case studies across Sub-Saharan Africa and humanitarian settings.\cite{IRENA2022,grafham2022,GPA2022}

These vulnerabilities are particularly acute in fragile off-grid contexts, such as rural villages, islands, or displacement settings, where energy access is not simply a matter of sustainability, but one of basic survival. Yet, the deployment of infrastructure and technologies is only the beginning. 
As systems are increasingly installed in remote areas, their inability to remain operational over time reveals a deeper and persistent issue: the lack of effective, on-site Operation and Maintenance (O\&M). Technology and financing, while necessary, are not sufficient. Without robust O\&M systems, local ownership, and cross-sector cooperation, even adequately funded projects can face premature failure. 
This fragility is particularly evident in displacement settings, where humanitarian actors often lack structured protocols to ensure the long-term availability of energy systems.\cite{IRENA2022,GPA2022,IRENA2023} The result is a systemic breakdown: one where the energy challenge shifts from deployment to long-term system availability.

In this perspective, we argue that ensuring the long-term availability of energy services, particularly through sustainable Operation and Maintenance (O\&M) systems, must become a core priority in future energy access strategies, especially in remote and fragile off-grid settings where the interplay of predictable and unforeseen factors heightens the risk of service disruption.

\section{Structural and contextual barriers}
In recent years, discussions around the United Nation's Sustainable Development Goal 7 (SDG 7) have focused heavily on sustainability, particularly the expansion of clean energy sources.\cite{SDG7} This focus is justified by the fact that decarbonizing the energy mix is essential to mitigate climate change and reduce environmental harm through lower emissions. Efforts have therefore concentrated on increasing the share of renewables especially in electricity generation, scaling up solar and wind technologies, and progressively phasing out fossil fuels.\cite{REN21} At the same time, attention has been given to the need for public acceptance and social legitimacy of these new technologies, especially in regions where energy transitions may affect livelihoods and local economies.

Yet, other critical aspects of energy access still need to be addressed. In particular, the idea of ensuring a stable and continuous energy supply, which is often referred to as energy security, deserves closer consideration. This includes building reliable infrastructure, ensuring resilience to disruptions, and managing energy systems in ways that can respond flexibly to fluctuations in demand and supply.\cite{IEA2024} These challenges are especially pressing where centralized energy systems are weak, intermittent, or entirely absent.

In these contexts, energy scarcity is not merely about the total absence of electricity but also about chronic unreliability, unaffordability, and volatility. According to the International Energy Agency, energy scarcity is defined as the lack of “uninterrupted availability of energy sources at an affordable price.” For many communities, particularly those in remote areas, this is the norm rather than the exception. Over 750 million people still lacked access to electricity in 2021, most of them in regions facing significant logistical, financial, and governance obstacles to clean energy delivery.\cite{IRENA2023} Electricity access in urban areas reaches 98\%, while in rural zones it drops to 85\%, and to just 56\% in the least-developed countries.\cite{IRENA2023} Without targeted and sustained interventions, hundreds of millions of people are likely to remain unelectrified beyond 2030. These statistics reflect the situation of communities in isolated rural settings, on islands or in mountains, and populations living in displacement settings such as refugee camps. Many of these groups rely on diesel generators that are both expensive and polluting, while transitions to renewables are often delayed by infrastructure limitations, market failures, or lack of institutional capacity. The communities most in need of off-grid electrification face steep barriers to implementation, as physical remoteness, weak infrastructure, and social marginalization compound the difficulty of energy transitions in these areas.\cite{IRENA2023} 

In response to these multifaceted challenges, decentralized energy systems, especially mini- and micro-grids, emerge as a strategic response. Unlike grid extension, which is often economically infeasible in low-density areas, mini- and micro-grids offer scalable, modular, and adaptable solutions. Today, around 51\% of mini-grids worldwide are solar or solar-hybrid, with others powered by hydro, fossil fuels, or wind.\cite{WorldBank2022} In humanitarian settings, decentralized models have already shown their potential: from the Azraq refugee camp in Jordan to projects across Sub-Saharan Africa, DRE deployment is demonstrating how clean energy can be delivered even under extreme constraints.\cite{grafham2022,IRENA2023,UgandaUNHCR,RwandaUNHCR,JordanUNHCR} Case studies from Uganda, Rwanda, Kenya, and Ethiopia have highlighted both the opportunities and limitations of deploying decentralized renewables in displacement and off-grid contexts, reinforcing the need for place-based, community-led approaches.\cite{UgandaUNHCR,RwandaUNHCR,JordanUNHCR}

\section{The hidden challenge: Operation and Maintenance}
This broader reading of SDG 7, which considers not only sustainability but also long-term system availability and reliability, invites a shift in focus: from simply delivering clean energy infrastructure to ensuring that such systems remain operational over time, especially in off-grid or underserved areas.\cite{SDG7} In these contexts, the issue is not just whether energy is clean, but whether it is consistently available, locally manageable, and resilient to shocks. Without long-term system availability, even the most sustainable systems fail to meet people’s basic needs.
Yet, a sole focus on infrastructure deployment has left structural barriers unaddressed. Despite improved commercial performance, the off-grid solar sector faces significant challenges.
According to the \textit{Off-Grid Solar Market Trends Report 2024}, 78\% of stakeholders cite a lack of technical skills as a primary barrier to scale, especially for maintaining systems in remote locations.\cite{OffGrid2024}
Inadequate funding for O\&M, particularly in public institutions such as schools and hospitals, frequently leads to premature system failure. Batteries often go unreplaced. As a result, donors are forced to reinvest in restoring neglected systems instead of expanding coverage.\cite{OffGrid2024} 

Operational fragility of DRE systems in remote regions is further amplified by environmental factors such as storms, dust, and extreme heat, which degrade performance over time.\cite{IRENA2022} In many cases, maintenance responsibility is delegated to end-users such as teachers or healthcare workers, who often lack the training to perform even basic repairs, resulting in a disproportionate distribution of operational tasks. This sentiment is echoed in the words of Raihan Elahi from the World Bank: "We expect that the doctors [...] will also cure solar systems and replace the batteries. […] Within two to three years, it stops [working], then new donors come in with new money to fund the same systems".\cite{IRENA2022}

This vulnerability affects energy users as well as local service providers. Many off-grid solar companies operate under pay-as-you-go (PAYG) business models, where revenue depends on the system’s proper functioning and the customer's ability to continue making payments.\cite{GOGLA} In usage-based PAYG schemes, payments are often interrupted when a system breaks down, making regular maintenance critical to sustaining income. 
When failures go unresolved, due to lack of technical training, spare parts, or accessible local support, the consequences are twofold.
On the customer side, despite improving short-term affordability, PAYG models often result in households paying more than the system’s upfront retail price without the assurance that it will remain functional throughout the full repayment period. This erodes trust and can deter long-term adoption, especially in the absence of proper maintenance. On the provider side, system downtime severely affects business viability, often extending return-on-investment periods and pushing many installers toward bankruptcy.

\section{Toward long-term system availability in remote off-grid contexts}
In Sub-Saharan Africa, O\&M responsibilities are often combined with the installer’s role, where the same entity acts as the Engineering, Procurement, and Construction (EPC) contractor, asset manager, and operator. While this model works in early-stage markets, it lacks long-term specialization and accountability, which limits the sustainability of energy systems.\cite{AfricaEd} To address these challenges, alternative models are emerging. These include third-party contractors dedicated solely to O\&M, hybrid approaches where local technicians are trained and remotely supported, and community-led micro-utilities embedded within local governance structures. Platform-based models also offer centralized monitoring and diagnostics, though interoperability and connectivity issues remain.\cite{AfricaEd} 

Recent recommendations highlight the need for private-sector-led O\&M models supported by donors, governments, and service  providers.\cite{IRENA2022} They emphasize building resilience from the outset of energy system deployment, through robust design, quality components, and preventive maintenance.\cite{IRENA2022} Plug-and-play solutions have been proposed to reduce reliance on highly skilled technicians and enable faster, decentralized maintenance. These systems empower community-level actors to perform routine tasks with minimal tools and training.\cite{IRENA2022} However, scalability is often hindered by limited interoperability, especially in complex humanitarian settings.\cite{Nature}

For standalone off-grid systems, particularly in remote and fragile contexts, best practices prioritize simplicity, cost-efficiency, and ease of maintenance with minimal equipment and training. Modular designs facilitate diagnostics and component replacement, while intuitive, visual documentation accommodates varying levels of literacy and technical expertise.\cite{AfricaEd} Monitoring tools should function offline and syncronize when possible, using open standards to ensure long-term system availability even if the original provider withdraws. Preventive maintenance following annual plans is essential, supported by a basic inventory of spare parts and pre-identified supply chains for more complex repairs.\cite{AfricaEd}

Local capacity building is critical. Embedding trained technicians within community structures supports both long-term system availability and operational resilience. Without embedded O\&M, the technical potential of energy systems cannot translate into sustained energy access.\cite{AfricaEd}

Ultimately, O\&M is a strategic function, not merely a support service. Especially in remote contexts, it determines whether electrification efforts endure or fade away. Embedding resilience from design through end-of-life is the only way to shift from temporary interventions toward long-term system availability. 
The pathway we propose focuses on making both preventive and corrective maintenance more user-friendly, delivering clear benefits to all stakeholders. This approach helps installers sustain their operations, extends the lifespan of energy systems, and, when supported by appropriate tools, fosters self-sufficiency and education within local communities.

\section{Conclusions}
Investments in off-grid systems are steadily increasing across underserved regions. Yet, without resilient maintenance strategies, many of these systems fail to deliver on their long-term promise. In this perspective, we have argued that long-term system availability in off-grid and remote contexts hinges not only on technology deployment, but on the capacity of local communities and operators to keep systems running after installation. Also, we have emphasized the importance of designing maintenance strategies that are accessible and practical, such as rugged, easy-to-use toolkits and offline mobile apps for preventive diagnostics and basic repairs, thus empowering local operators, especially in low-connectivity, resource-scarce environments, to maintain solar, battery, and hybrid energy systems independently. This user-friendly approach helps installers sustain their services, keeps systems functional when external support is unavailable, and promotes self-sufficiency through skills transfer and learning-by-doing. 

Ultimately, embedding maintenance into the daily practice of decentralized energy systems is not a peripheral task. It is a structural requirement for achieving lasting impact. In doing so, our proposed shift in focus directly contributes to SDG 7 by reinforcing clean and reliable energy access. It also supports SDGs 3 and 4, by enabling stable power supply in health and educational facilities and fostering skills-based learning. Furthermore, by promoting local employment and supporting inclusive economic growth alongside resilient, clean infrastructure, it addresses key targets of SDGs 8 and 9.

We hope that this proposed paradigm shift, if adopted and scaled, will provide a replicable pathway to bridge the long-term system availability gap in off-grid contexts, shifting the focus from infrastructure deployment to long-term resilience.

\section*{Acknowledgement}
We gratefully acknowledge the Innovation4Change educational programme, powered by Collège des Ingénieurs, for providing the intellectual and collaborative framework in which this work was developed. In particular, we thank Alissa Bauer and Erika Vaniglia for their guidance and support. We also acknowledge Elettrici senza Frontiere Italia for the insightful discussions that helped refine the practical dimension of our approach.

\end{document}